\documentclass[aps,showpacs,twocolumn,superscriptaddress,floatfix]{revtex4}
\usepackage{graphicx}
\usepackage{dcolumn}
\usepackage{bm}
\usepackage{amssymb}
\usepackage{amsmath}
\usepackage{amsfonts}
\usepackage{amssymb}
\usepackage{color}

\begin{document}

\title{Arbitrary Rotation of a Single Spinwave Qubit in an Atomic-Ensemble Quantum Memory}
\author{Jun Rui}
\affiliation{Hefei National Laboratory for Physical Sciences at Microscale and
Departmentof Modern Physics, University of Science and Technology of
China,Hefei,Anhui 230026, China}
\affiliation{CAS Center for Excellence and Synergetic Innovation Center in Quantum Information and Quantum Physics, University of Science and Technology of China, Hefei, Anhui 230026, China}
\author{Yan Jiang}
\affiliation{Hefei National Laboratory for Physical Sciences at Microscale and
Departmentof Modern Physics, University of Science and Technology of
China,Hefei,Anhui 230026, China}
\affiliation{CAS Center for Excellence and Synergetic Innovation Center in Quantum Information and Quantum Physics, University of Science and Technology of China, Hefei, Anhui 230026, China}
\author{Bo Zhao}
\affiliation{Hefei National Laboratory for Physical Sciences at Microscale and
Departmentof Modern Physics, University of Science and Technology of
China,Hefei,Anhui 230026, China}
\affiliation{CAS Center for Excellence and Synergetic Innovation Center in Quantum Information and Quantum Physics, University of Science and Technology of China, Hefei, Anhui 230026, China}
\author{Xiao-Hui Bao}
\affiliation{Hefei National Laboratory for Physical Sciences at Microscale and
Departmentof Modern Physics, University of Science and Technology of
China,Hefei,Anhui 230026, China}
\affiliation{CAS Center for Excellence and Synergetic Innovation Center in Quantum Information and Quantum Physics, University of Science and Technology of China, Hefei, Anhui 230026, China}
\author{Jian-Wei Pan}
\affiliation{Hefei National Laboratory for Physical Sciences at Microscale and
Departmentof Modern Physics, University of Science and Technology of
China,Hefei,Anhui 230026, China}
\affiliation{CAS Center for Excellence and Synergetic Innovation Center in Quantum Information and Quantum Physics, University of Science and Technology of China, Hefei, Anhui 230026, China}
\date{\today}

\begin{abstract}
We report the first experimental realization of single-qubit manipulation for single spinwaves stored in an atomic ensemble quantum memory. In order to have high-fidelity gate operations, we make use of stimulated Raman transition and controlled Lamor precession jointly. We characterize the gate performances with quantum state tomography and quantum process tomography, both of which imply that high-fidelity operations have been achieved. Our work complements the experimental toolbox of atomic-ensemble quantum memories by adding the capability of single-qubit manipulation, thus may have important applications in future scalable quantum networks.
\end{abstract}

\maketitle

Many physical systems have been experimentally studied for quantum information applications \cite{Binosi2013}. Among them, atomic ensembles \cite{Sangouard2011} are mainly famous for the capability of long-term storage of quantum states and the capability of efficient interaction with single-photons. In recent years, significant achievements on memory lifetime \cite{Radnaev2010,Dudin2013} and efficiency \cite{Simon2007,Bao2012} have been made. Nevertheless, direct manipulation of single qubits in an atomic-ensemble quantum memory has not been realized so far, which limits further applications of atomic-ensemble quantum memories. For instance, in order to teleport a qubit from one ensemble to another \cite{Bao2012} or entangle remote atomic ensembles \cite{Yuan2008} through interfering independent photons, random single-qubit manipulations on one ensemble are required in order to restore the target states conditioned on different Bell state measurement results. However, due to the incapability of qubit manipulation, in all previous experiments \cite{Chen2008,Yuan2008,Bao2012pnas}, one has to convert the spinwave states into single-photon states and apply the qubit rotations on photons instead. Inefficiency is the main drawback for this alternative way. For large-scale applications, the overall efficiency can be extremely low. If the spinwave states can be directly manipulated and detected in the atomic ensemble quantum memories, the efficiency will be largely enhanced.

Arbitrary single-qubit rotation is an essential element in quantum information science, and has been successfully implemented in single emitter systems, such as single neutral atoms \cite{Specht2011}, single ions \cite{Stute2013} and single quantum dot \cite{DeGreve2012b,Gao2012} and single NV center \cite{Togan2010, Maurer2012}. However, coherently manipulate the spinwave qubit in atomic ensembles remains challenging. For single-emitter systems, single-qubit manipulation can be easily realized by addressing the single emitter using microwaves or radio frequency fields. Nevertheless, qubit manipulation for a single spinwave qubit requires coherently manipulate all the atoms simultaneously, since a single spinwave is a collective superposition of all the atoms with only one excitation \cite{Duan2001, Fleischhauer2005}. The main difficulty of manipulating such a single spinwave lies in how to manipulate a single collective excitation mode precisely and coherently without influencing (depopulating or exciting) the rest majority atoms. Even a single indeliberate excitation from the majority of atoms could possibly ruin a stored single spinwave. The manipulation process has also to preserve the spinwave wave-vector since its amplitude and direction determines the mode direction for the retrieval photons. Preliminary study on manipulation of classical spinwaves has been performed previously by H. Wang et al. Either by manipulating the majority atoms \cite{Li2011} or the excited atoms \cite{Xu2013f}, they observed the flopping between two spinwave modes. Nevertheless, their experiments still lie in the classical regime, and number of excitation is on the order of $\sim$ $10^6$ which is far more larger than a single spinwave.

In this paper, we report the first experiment of single-qubit manipulation for single spinwaves in an atomic ensemble quantum memory. High-fidelity operation is achieved by making use of stimulated Raman transition and controlled Larmor procession jointly. An arbitrary single spinwave state is heraldedly prepared through the process of spontaneous Raman scattering and making projective measurement on the scattered single photons. For the realization of an arbitrary single-qubit operation, the capabilities of rotating an arbitrary angle along two orthogonal axes in the Bloch sphere are required. For the rotation along the axis in the direction of poles, we make use of controlled Larmor procession. For the rotation along an axis in the equatorial plane, we make use of stimulated Raman transitions. Experimental results are characterized with quantum state tomography and quantum process tomography.

\begin{figure}[htb]
\includegraphics[width=\columnwidth]{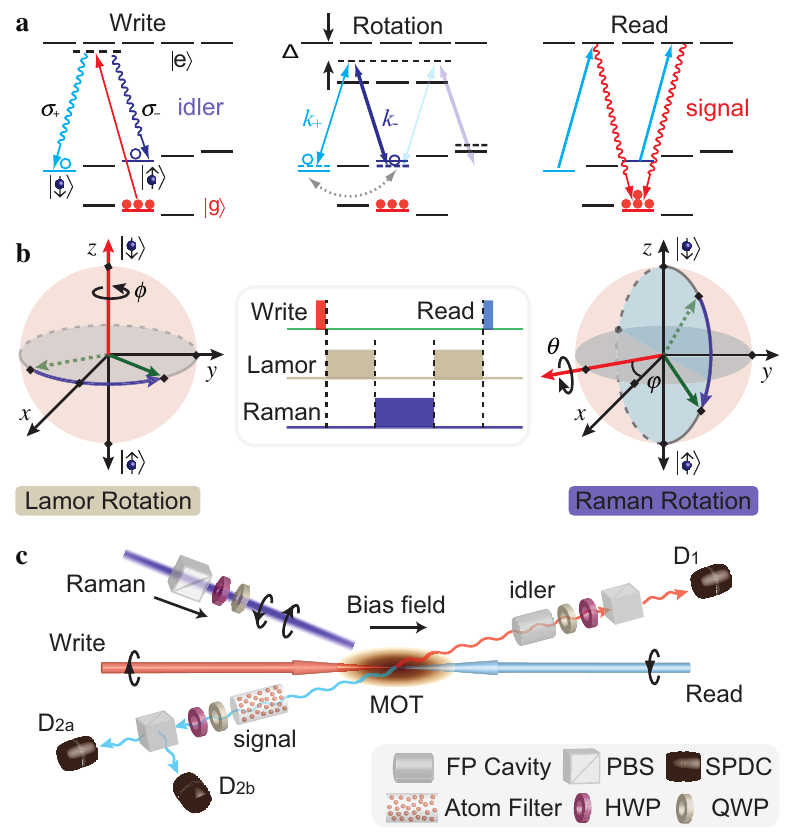}
\caption{Experiment scheme for arbitrary rotations of spinwave qubit. \textbf{a}, Energy levels: The write process projects the spinwave onto a superposition state between $|s_{\downarrow}\rangle$ and$|s_{\uparrow}\rangle$, by measuring the idler photon along a specific polarization state.  Then a series of rotations between two basis states are carried out. At last, the spinwave qubit state is verified by converting it back to a signal photon during the read process. The levels chosen here are designed to provide almost identical retrieval efficiencies for both spinwave states, since the transition strengths of $|F'=2, m_{f}=-1/+1\rangle \to |g\rangle$ are the same. \textbf{b}, Two elemental rotations in the Bloch sphere. One is the rotation along z axis, which is due to different Larmor precession frequencies of these two spin states. The other is along an axis in the x-y plane, where the orientation angle $\varphi$ is determined by the relative phase between two Raman beams. \textbf{c}, Experimental Layout. The write beam has a waist of 150 $\mu$m, a power of 1.5 $\mu$W and a detuning of -10 MHz. The idler photon is collected in the direction with an angle of $1.5^{\circ}$ relative to the write beam, and has a mode waist of 100 $\mu$m.
It is further frequency-filtered with a Fabry-Perot cavity. While the signal photon is filtered by a Rubidium vapor filter to avoid back reflections into the idler channal during write process. The polarization of a single Raman beam is prepared to provide both helicities for stimulated Raman transitions between $|s_{\downarrow \setminus \uparrow }\rangle$ . }
\label{fig1}
\end{figure}

Our experimental scheme is shown in Fig. \ref{fig1}. An ensemble of $^{87}$Rb atoms are captured using a standard magneto-optical trap (MOT). With 2 ms polarization gradient cooling, the atomic ensemble is cooled to a temperature of about 10 $\mu$K. All the atoms are initially pumped to the state of $|g\rangle \equiv |F=1, m_{f}=0\rangle$. To create a single spinwave exication \cite{Duan2001}, as shown in Fig. \ref{fig1}(a), a $\sigma^-$ polarized write pulse is applied to couple the transition of $|g\rangle\rightarrow|e\rangle$ with $|e\rangle \equiv |F'=2, m_{F}=-1\rangle$ weakly. Raman scattered $\sigma^+$ and $\sigma^-$ idler photons are selected through spacial and frequency filtering. A $\sigma^+$ idler photon heralds the creation of a single excitation in $|\downarrow\rangle\equiv|F=2, m_{f}=2\rangle$, while a $\sigma^-$ idler photon heralds the creation of a different single excitation in $|\uparrow\rangle \equiv |F=2,m_{f}=0\rangle$. For each write pulse, the heralding probability is $3\times 10^{-3}$, which guarantees that probability for creating multiple ($\geq$ 2) excitations is negligibly small. Interference between these two channels gives rise to the entanglement between an idler photon and an atomic spinwave \cite{Matsukevich2005,Xu2012a} in the form of
\begin{equation*}
|\Psi_{AP}\rangle =\sqrt{2/5} |s_{\downarrow}\rangle |\sigma^{+}\rangle - \sqrt{3/5} |s_{\uparrow}\rangle |\sigma^{-}\rangle,
\end{equation*}
where the amplitude and relative phase are determined by the atomic transitions and the collective spin states are defined as $|s_{\downarrow}\rangle=\Sigma_{j}e^{i\vec{k}_{s} \cdot \vec{r}_{j} }|g_{1}...\downarrow_{j}...g_{N}\rangle$ and $|s_{\uparrow}\rangle=\Sigma_{j}e^{i\vec{k}_{s} \cdot \vec{r}_{j} }|g_{1}...\uparrow_{j}...g_{N}\rangle$, with $\vec{k}_{s}=\vec{k}_{w}-\vec{k}_{i}$.
Once the idler photon is measured under specific polarizations, the spinwave will be projected to the corresponding states.

In order to prepare an arbitrary single spinwave state of $|\psi_{s}\rangle=\cos\theta |s_{\downarrow}\rangle + \sin\theta e^{i\phi} |s_{\uparrow}\rangle$, one just needs to project the idler photon onto a corresponding state of $|\psi_{i}\rangle=(\sqrt{3/5}\cos \theta - \sqrt{2/5}\sin \theta e^{-i\phi}) |H\rangle +i (\sqrt{3/5}\cos \theta + \sqrt{2/5}\sin \theta e^{-i\phi}) |V\rangle$, where a normalizing factor is omitted. The projection of idler photon can be done through standard linear-optics method~\cite{Englert01}. Before performing qubit rotations on this spinwave state, we first charactize the state preparation fidelities. The spinwave state is measured by applying a strong read pulse which converts $|s_{\downarrow}\rangle$ to a $\sigma^{-}$ polarized signal photon and converts $|s_{\uparrow}\rangle$ to $\sigma^{+}$ polarized signal photons. Arbitrary single-qubit measurement is realized through measuring the signal photon in arbitrary polarization bases, which is also one of the highlights of our system. In our experiment, we select six spinwave states to prepare: $|s_{\downarrow}\rangle$, $|s_{\uparrow}\rangle$, $|s_{D}\rangle$, $|s_{A}\rangle$, $|s_{R}\rangle$ and  $|s_{L}\rangle$, with $|s_{D/A}\rangle=1/\sqrt{2}( |s_{\downarrow}\rangle \pm |s_{\uparrow}\rangle )$ and $|s_{R/L}\rangle=1/\sqrt{2}( |s_{\downarrow}\rangle \pm i|s_{\uparrow}\rangle )$. We make use of quantum state tomography and the maximum likelihood method\cite{James01} to characterize the prepared states, and calculate the fidelities in comparison with the target states. Results are shown in Tab.\ref{tab1}, and an average fidelity of $97.2\%$ is achieved. Standard errors for each state are calculated by poissonian sampling of the measured counts for 200 times.

\begin{table}[htb]
\includegraphics[width=\columnwidth]{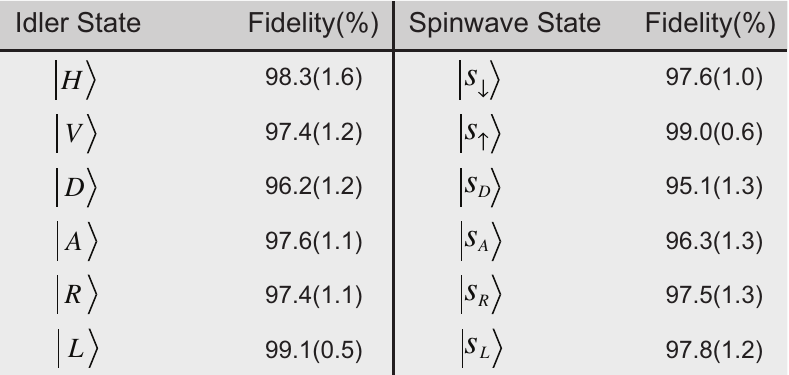}
\caption{Fidelities of state preparations. (Left) Fidelities between measured and expected signal photon states for different idler photon polarizations. Fidelities are relatively higher due to the polarization needed is more accurate to prepare. (Right) Fidelities between measured and expected signal photon states for different spinwave states. }
\label{tab1}
\end{table}

In order to realize arbitrary single-qubit manipulation, the ability of rotation along two orthogonal axes in the Bloch sphere are required. As shown in Fig. \ref{fig1}(b), we first make use of the Larmor procession process to realize the rotation of $R_{z}(\phi)$ which corresponds to rotation along an axis in the direction of poles. As the magnetic moment of the $|s_{\uparrow}\rangle$ state is zero, the Larmor phase of this state is constant during the experiment. While for the $|s_{\downarrow}\rangle$ state, its magnetic moment precesses around the bias magnetic field, with an angluar frequency of $\omega_{L}=2\pi B_{0}\times 1.4$ MHz/Guass. Thus the relative phase between $|s_{\downarrow}/s_{\uparrow}\rangle$ will evolve as $\phi(t)=\omega_{L}t$, which effectively is a rotation around z-axis in Bloch sphere. A specific rotation angle can be realized by changing the duration of the Larmor procession. With the bias field applied, we measured the Larmor frequency to be $\omega_{L}=2\pi \times 180$kHz. The rotation matrix of Larmor operation is,
\begin{equation*}
R_{z}(t_{L})=\left[\begin{array}{cc}
e^{i \Omega_{L} t_{L}}  & 0 \\
0  & 1
\end{array}\right],
\end{equation*}
where $t_{L}$ is the duration of Larmor operation.

We make use of stimulated Raman transition to realize the rotation along an axis in the equatorial plane. The detailed scheme and experimental arrangement are shown in Fig. \ref{fig1}. First, we notice that if both Raman beams are red (or blue) detuned relative to the $|F'=2\rangle$ and $|F'=1 \rangle$ hyperfine states of the D$1$ line, destructive interference between these two Raman transition channels will occur. To avoid this, we set the frequency of the Raman beams to be at the middle of these two states, as shown in Fig. \ref{fig1}(a), where the detuning is $\Delta \approx$ 408 MHz. Thus these two Raman transitions constructively interfere with each other, and lead to faster rotation operations with limited laser power. Second, it's found that when the Raman beams are applied, they couple not only the transition of $|s_{\uparrow}\rangle \leftrightarrow |s_{\downarrow}\rangle$ but also an unexpected transition of $|s_{\uparrow}\rangle \leftrightarrow |s_{aux}\rangle \equiv |F=2, m_{F}=+2\rangle$. As the state $|s_{aux}\rangle$ cannot be converted to signal photons in the read process, we should minimize the population leakage into this state, and restrict the Rabi flopping to occur mainly between $|s_{\downarrow}\rangle$ and $|s_{\uparrow}\rangle$.

The Raman beam in our setup has a total power of about 7 mW, with a waist of 1.9 mm. The polarization state of the beam is adjusted to be $|\psi_{R}\rangle=\sqrt{1/7} |\sigma^+\rangle + e^{i
\varphi_{R}} \sqrt{6/7}|\sigma^-\rangle$, where $\sigma^+/\sigma^-$ terms serve as the $k^+/k^-$ Raman beams in Fig. \ref{fig1}(a) respectively, and $\varphi_{R}$ is the relative phase between these two terms. With such a polarization, AC Starks shifts due to the Raman beams of $|s_{\downarrow}\rangle$, $|s_{\uparrow}\rangle$ and $|s_{aux}\rangle$ states are calculated to be about +40 kHz, -140 kHz and +240 kHz, respectively. Thus the frequency splitting between $|s_{\uparrow}\rangle$ and $|s_{\downarrow}\rangle$ induced by AC Starts effect is -180 kHz, which just cancels the Zeeman splitting (+180 kHz) between these two states. However, the overall frequency splitting between $|s_{aux}\rangle$ and $|s_{\uparrow}\rangle$ are shifted to an even larger value of 560 kHz. The theoretical two-photon Rabi frequencies of $|s_{\uparrow}\rangle \leftrightarrow |s_{\downarrow}\rangle$ and $|s_{\uparrow}\rangle \leftrightarrow |s_{aux}\rangle$ are both 240 kHz. As the two-photon detuning of the $|s_{\uparrow}\rangle \leftrightarrow |s_{aux}\rangle$ transition is significantly larger than the Raman Rabi frequency, the maximum population leakage onto $|s_{aux}\rangle$ during Rabi flopping is estimated to be $\sim1\%$. Thus we can restrict the population mainly flops between our target spinwave states of $|s_{\uparrow}\rangle$ and $|s_{\downarrow}\rangle$.

Then we can successfully create a pure rotation along a definite axis in the x-y plane of the Bloch sphere when a Raman pulse is applied. Note the angle $\varphi$ between the Raman rotation axis and x axis is determined by the relative phase $\varphi_{R}$ between $k^+$ and $k^-$ Raman components. An additional merit of our scheme is that this phase can be precisely controlled and it's stable in itself, excluding other electronic phase stabilizing controls. The rotation matrix of Raman operation can be written as,

\begin{equation*}
R_{\vec{n}}(t_{R})=\left[\begin{array}{cc}
\cos{\phi(t_{R})} & -ie^{i\varphi_{R}}\sin{\phi(t_{R})} \\
-ie^{-i\varphi_{R}}\sin{\phi(t_{R})}  & \cos{\phi(t_{R})}
\end{array}\right],
\end{equation*}
where $\phi(t_{R})=\Omega_{R}t_{R}/2$, and $t_{R}$ is the duration of applied Raman pulse. By choosing $\varphi_{R}=0$, rotation along x-axis can be realized. While with $\varphi_{R}=-\pi/2$, the rotation is along y-axis, as shown in Fig. \ref{fig1}(b). With Raman beam paramters stated above, the measured Raman Rabi frequency is about 190 kHz. The deviation with theoretical predications may be due to the imperfect interference between two Raman transition channels.

\begin{figure}[htb]
\includegraphics[width=\columnwidth]{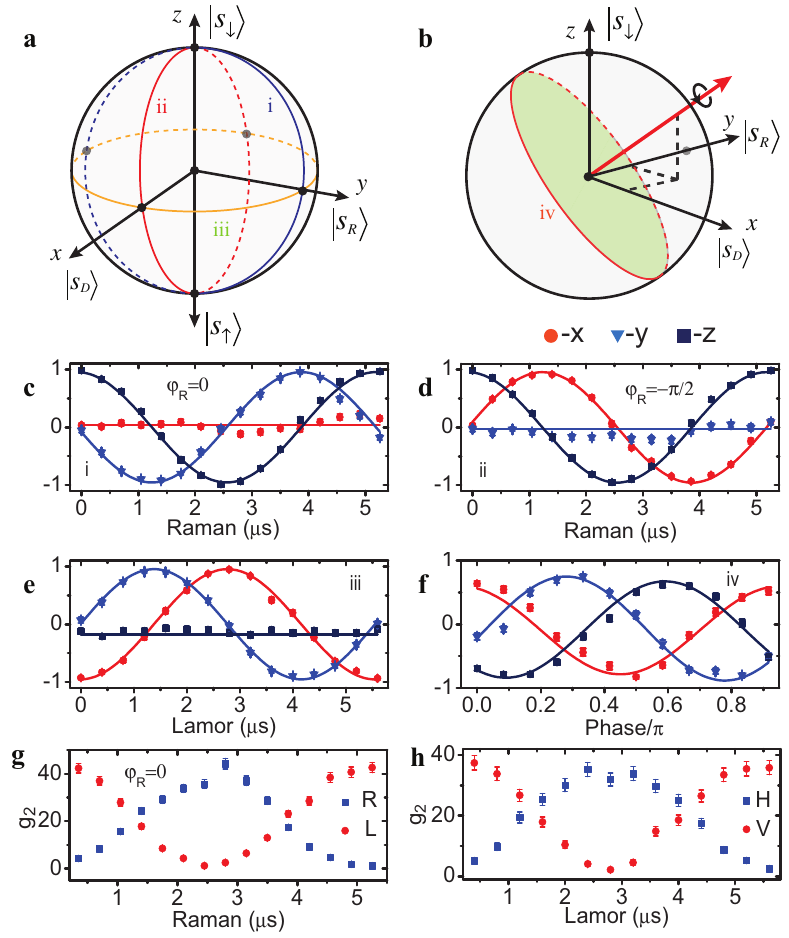}
\caption{Arbitrary rotations of the Spinwave qubit. \textbf{a-b}, Illustratons of state rotations in the Bloch sphere. \textbf{a}, rotations along x-, y-, z-axis. \textbf{b}, special rotation along axis $\vec{n}=\{ \frac{1}{\sqrt{3}},\frac{1}{\sqrt{3}},\frac{1}{\sqrt{3}} \}$, and the rotation plane crosses zero point. \textbf{c-f}, Expreimental results of rotations, dotted points are measured values of each axis, and compared with theoretical curves. \textbf{c}, $R_{x}$ rotations realized by Raman operations with $\varphi_{R}=0$. \textbf{d}, $R_{y}$ rotations by Raman operations with $\varphi_{R}=-\pi/2$. \textbf{e}, $R_{z}$ rotations by Larmor operations. \textbf{f}, $R_{\vec{n}}$ rotations by combinations of $R_{z}$ and $R_{x}$ rotations. \textbf{g-h}, non-classical correlations between idler and signal photons after operations for different states of signal photons. \textbf{g}, the initial spin state is on $|s_{\downarrow}\rangle$, Raman rotations with $\varphi_{R}=0$ are applied. \textbf{h}, the initial spin state is approximately on $|s_{A}\rangle$ with the idler photon projected along $|H\rangle$ state. }
\label{fig2}
\end{figure}

We first evaluate the gate performances with quantum state tomography ~\cite{James01} by measuring the quantum state of the spinwave after a single-qubit operation and compare it with the ideal target state using $F(\rho_{1}, \rho_{2}) \equiv \{ $tr$[ (\sqrt{\rho_{1}} \rho_{2} \sqrt{\rho_{1}} )^{\frac{1}{2}} ] \}^{2}$. Rotations of $R_{x}, R_{y}, R_{z}$, are carried out for a series of rotation angles as shown in Fig. \ref{fig2}(a). For Raman $R_{x}, R_{y}$ rotations, the prepared initial spin state is $|s_{\downarrow}\rangle$. And for Larmor $R_{z}$ rotation, the initial spin state is projected by choosing the state of idler photon to be $|H\rangle$. In Fig. \ref{fig2}(c $\sim$ e), Stokes parameters are presented for each rotation, and are compared with theoretical curves which fit well. The state fidelities of $R_{x}, R_{y}, R_{z}$ operations averaged for all rotation angles are $98.6(0.6)\%, 98.8(0.5)\%$ and $99.1(0.3)\%$ respectively. Note here the expected state is generated by applying ideal operations on the measured density matrix of initial states, which takes account of imperfections of state preparations and diagnose the rotation operations exclusively. To show how well the quantum nature is preserved after operations, non-classical correlations between idler and signal photons are shown in Fig. \ref{fig2}(g) and(h), where the visibilities are both well above 6. Moreover an arbitrary rotation can be decomposed as $R_{\vec{n}}(\phi)=R_{z}(\alpha)R_{y}(\beta)R_{z}(\gamma)e^{i\delta}$ \cite{nielsen00} by combining $R_{y}$ and $R_{z}$ operations. As an example, we choose to realize $R_{\vec{n}}(\phi)=\exp(-i\vec{n}\cdot\vec{\sigma} \phi )$ with $\vec{n}=(\frac{1}{\sqrt{3}}, \frac{1}{\sqrt{3}},\frac{1}{\sqrt{3}})$, which is illustrated in Fig. \ref{fig2}(b). To make the rotation plane meet with zero point, we prepare the initial spin state on $|\psi_{s}\rangle=\cos\frac{3}{8}\pi |s_{\downarrow}\rangle+\sin\frac{3}{8}\pi |s_{\uparrow}\rangle$, with a state fidelity of $96.6(1.7)\%$. Then a series of rotation operations are performed on the initial state,  with roation angles $\phi=0, \frac{1}{12}\pi,..., \frac{11}{12}\pi$. For each rotation, it's decomposed into values of $\alpha, \beta, \gamma$, with the global phase term $\delta$ neglected. The result is shown in Fig. \ref{fig2}(f), with an average state fidelity of $98.3(0.8)\%$.

In order to give more complete characterization for our single-qubit gate, we make use of quantum process tomography \cite{nielsen00}. An arbitrary single-qubit operation on an input state $\rho_{in}$ can be fully described by a process matrix $\chi$, which is defined as $\rho_{out}=\sum_{i,j=0}^{3} \chi_{i,j} \sigma_{i} \rho_{in} \sigma_{j}^{\dagger} $, where $\sigma_{i}$ are Pauli matrices with $\sigma_{0}=I, \sigma_{1}=\sigma_{x}, \sigma_{2}=\sigma_{y}, \sigma_{3}=\sigma_{z}$. The distance between paractical and ideal operations can be characterized by process fidelity, $F_{proc}=$tr$[\chi_{1}\chi_{2}]$. In order to measure the process matrix for a single-qubit operation, we select six different spinwave states, which are listed in Tab.\ref{tab1}, as input states and carry out quantum state tomography for each output state. In order to get a physical process matrix we make use of the maximal likelihood method \cite{OBrien04}. In our experiment quantum process tomography is measured for Pauli operations ($\sigma_x$, $\sigma_y$ and $\sigma_z$) and the Hadamard gate, with the results shown in Tab.\ref{tab2}. The process fidelity averaged for these four operations is calculated to be 94.7(7)\%. With the process fidelities, one can estimate the average state fidelity $F_{ave}$ of a process, which is defined as the average of state fidelities between input and output states, where input states are randomly selected from the Bloch sphere. It's shown that, the average fidelity of a process, can be simplified to evaluate the mean state fidelities of six pure input states locating at the cardinal points \cite{Bowdrey2002}. The average fidelity is related with the process fidelity by, $F_{ave}=\frac{dF_{proc}+1}{d+1}$, where $d$ is the dimensionality of the system and $d=2$ in our case\cite{Nielsen2002}. Both measured average fidelity $F_{ave}$ and average fidelity $F_{ave}^{th}$ derived from $F_{proc}$ are also listed in Tab.\ref{tab2}. We note that $F_{ave}$ is slightly higher than $F_{ave}^{th}$, which is mainly due to the fact that the prepared initial states are not pure and thus a different definition of $F(\rho_{1},\rho_{2})$ is adopted. Besides, we notice that, the constraint of completely positive mapping on the fitted process matrixes, also reduces the calculated process fidelities, thus further lowering down the $F_{ave}^{th}$ values in Tab.\ref{tab2}.

\begin{table}[htb]
\includegraphics[width=\columnwidth]{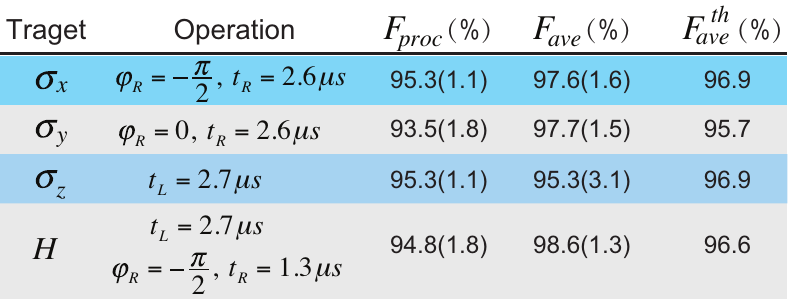}
\caption{Process Fidelity of Pauli and Hadamard Operations. Pauli operations and Hadamard gate are realized by choosing appropriate Raman and Larmor parameters. The measured process matrices by maximum likelihood method  are compared with matrices of ideal operations, to determine the process fidelities. Average state fidelities of each operation from measured results are also shown here.  }
\label{tab2}
\end{table}

To summarize, by making use of stimulated Raman transition and controlled Larmor procession, we have implemented single-qubit operations for single spinwaves in an atomic ensemble quantum memory for the first time. We have made use of quantum state tomography to measure the target states for arbitrary rotations. Average state fidelity for $R_x$, $R_y$ and $R_z$ rotations is measured to be 98.8(3)\%. We have also adopted the method of quantum process tomography to characterize the Pauli operations and the Hadamard gate. Average process fidelity is measured to be 94.7(7)\%. By making use of Raman beams with better intensity homogeneity and actively controlling the ambient magnetic field, even higher fidelities can be got. Our work enriches the experimental toolbox of harnessing atomic ensembles for high-performance quantum memories, thus could possibly have lots of applications in future scalable quantum networks \cite{Kimble2008a}.

This work was supported by the National Natural Science Foundation of China, National Fundamental Research Program of China (under Grant No. 2011CB921300), and the Chinese Academy of Sciences. B.Z. and X.-H.B. acknowledge support from the Youth Qianren Program.

\bibliography{qubitrefs}

\end{document}